\documentclass[12pt,a4paper,english,superscriptaddress]{revtex4}
\usepackage{graphicx}
\usepackage{graphics}
\usepackage{amsmath}
\usepackage{dcolumn}
\usepackage{amssymb}
\usepackage{bm}
\usepackage[latin1]{inputenc}
\newcommand{\be}{\begin{equation}}
\newcommand{\ee}{\end{equation}}
\newcommand{\bea}{\begin{eqnarray}}
\newcommand{\eea}{\end{eqnarray}}

\begin{document}
\title{On the effective potential in higher-derivative superfield theories}

\author{M. Gomes}
\email{mgomes@fma.if.usp.br}
\affiliation{Instituto de F\'\i sica, Universidade de S\~ao Paulo\\
Caixa Postal 66318, 05315-970, S\~ao Paulo, SP, Brazil}

\author{J. R. Nascimento} 
\email{jroberto@fisica.ufpb.br}
\affiliation{Departamento de F\'{\i}sica, Universidade Federal da Para\'{\i}ba\\
 Caixa Postal 5008, 58051-970, Jo\~ao Pessoa, Para\'{\i}ba, Brazil}

\author{A. Yu. Petrov}
\email{petrov@fisica.ufpb.br}
\affiliation{Departamento de F\'{\i}sica, Universidade Federal da Para\'{\i}ba\\
 Caixa Postal 5008, 58051-970, Jo\~ao Pessoa, Para\'{\i}ba, Brazil}

\author{A. J. da Silva}
\email{ajsilva@fma.if.usp.br}
\affiliation{Instituto de F\'\i sica, Universidade de S\~ao Paulo\\
Caixa Postal 66318, 05315-970, S\~ao Paulo, SP, Brazil}

\begin{abstract}
We study the one-loop quantum corrections for 
higher-derivative superfield theories, generalizing the approach for
calculating the superfield effective potential. In particular, we calculate the effective potential for two versions of higher-derivative chiral superfield models. We point out that the equivalence of the higher-derivative theory for the chiral superfield and the one without higher derivatives but with an extended number of chiral superfields occurs only when the mass term is contained in the general Lagrangian. The presence of divergences can be taken as an indication of that equivalence.
\end{abstract}
\maketitle

\section{Introduction}

The study of higher-derivative field theories has a long
story.  In supersymetric models, the
higher-derivative regularization method was proposed in
\cite{Ili}.  In the context of gravity higher derivatives were introduced in 
\cite{stelle}  where it was shown that the
presence  of higher derivatives  greatly improves the
renormalization properties of field theories.  Further,
higher-derivative modifications of the gravity action were shown to
arise due to the presence of the conformal anomaly of matter fields in
curved space \cite{AM}.The superfield generalization of this concept,
based on the study of the supertrace anomaly of matter superfield in
the curved superspace \cite{BK}, was carried out in  \cite{bp1}.
There, the higher-derivative action for the conformal sector (dilaton)
of the $N=1$ superfield supergravity, composed by the usual
supergravity action in the conformal sector plus an additive term,
generating the superconformal anomaly, was formulated. In the papers
\cite{bp2} the effective action for this theory was studied in
detail, and the superfield approach to the study of the effective
potential, earlier developed in \cite{efpot}, (see \cite{ep3d} for its
three-dimensional generalization) was successfully generalized for the 
higher-derivative theories. 

Actually, the higher-derivative field theories are studied in
different contexts, including different gravity modifications which
are intensively applied to obtain the cosmic acceleration \cite{Dol},
and the Horava model of  gravity \cite{Horava}. In the context of
supersymmetry, the interest in the higher-derivative superfield
theories was recently recovered due to the paper \cite{Ant}, where the
equivalence of the higher-derivative supersymmetric theories and the
ones with greater number of  superfields
was shown on the tree level (we notice that this idea can be interpreted
as a reminiscence of the method for constructing  an
effective action for  light superfields from a  theory involving light
and  heavy superfields
\cite{BCP}; see also \cite{bp3} for the study of the superfield
effective action in a generic case without higher
derivatives). Therefore, a natural question is whether such a mapping
between higher-derivative superfield theories and superfield theories
with an extended number of superfields is maintained on the
perturbative level. Another interesting problem is whether the
decoupling of the heavy states observed in \cite{BCP} occurs as well
in the higher-derivative theories. It is clear that to study these
problems one should first describe the general properties of 
effective 
actions in  higher-derivative superfield theories.

This paper has the following organization. First, we describe the general
structure of the one-loop effective action in higher-derivative
superfield theories. Second, we develop a procedure for the one-loop
calculation of the superfield effective potential in the
higher-derivative superfield theory, which turns out to imply in
different results for different ways of introducting the mass 
in the theory. 

\section{Effective action in higher-derivative superfield theories:
  general  approach}

Let us start with a following example of a higher-derivative superfield theory:
\bea
\label{firstex}
S[\Phi,\bar{\Phi}]=\int d^8z \Phi\Box\bar{\Phi}+(\int d^6z W(\Phi)+h.c.).
\eea
Here $\Phi$ is a chiral superfield, and $W(\Phi)$ is an  arbitrary
function. Within this paper we follow notations and conventions of
\cite{BK0}. A particular form of this action was studied in \cite{bp2}.
This action, being reduced to the components, contains fourth order in space-time derivatives (cf. \cite{bp1}).
We will refer to this theory as  the minimal higher-derivative
theory, since in this case, similarly to the minimal theory in
\cite{BCP}, 
all couplings are concentrated in the superpotential sector.

The effective action $\Gamma[\Phi,\bar{\Phi}]$, as usual, can be represented as a generating functional of the one-particle-irreducible vertex Green functions:
\bea
\label{def}
e^{i\Gamma[\Phi,\bar{\Phi}]}=\int D\phi D\bar{\phi} \exp(iS[\Phi+\phi,\bar{\Phi}+\bar{\phi}])|_{1PI}.
\eea
Here the $\Phi,\bar{\Phi}$ are background (classical) fields and $\phi,\bar{\phi}$ are  quantum fields.
Following \cite{BK0}, we can represent the structure of the effective action in this theory as
\bea
\Gamma[\Phi,\bar{\Phi}]=\int d^8z {\cal L}(\Phi,\bar{\Phi})+(\int d^6z {\cal L}_c(\Phi)+h.c.),
\eea
where ${\cal L}(\Phi,\bar{\Phi})$ is called general effective Lagrangian, which depends on superfields $\Phi,\bar{\Phi}$ and their derivatives, and ${\cal L}_c$ is called chiral effective Lagrangian which depends only on the chiral superfield $\Phi$ and its space-time derivatives, i.e. it is a chiral superfield itself.

To obtain the one-loop effective action, one should expand the right-hand side of the equation (\ref{def}) up to the second order in the quantum superfields $\phi$, $\bar{\phi}$ (cf. \cite{BO}). As a result, the one-loop effective action is defined from the expression:
\bea
\label{def1}
e^{i\Gamma^{(1)}[\Phi,\bar{\Phi}]}=\int D\phi D\bar{\phi} \exp(i[\int d^8z \phi\Box\bar{\phi}+(\frac{1}{2}\int d^6z W^{\prime\prime}(\Phi)\phi^2+h.c.)]),
\eea
which yields the following form:
\bea
\Gamma^{(1)}[\Phi,\bar{\Phi}]=\frac{i}{2}{\rm Tr}\ln\left(
\begin{array}{cc}
W^{\prime\prime}& -\Box\frac{\bar{D}^2}{4}\\
-\Box\frac{D^2}{4} &\bar{W}^{\prime\prime}
\end{array}
\right).
\eea
The elements of this matrix are defined in different subspaces of the
superspace and mix the chiralities. Therefore, the straightforward
calculating of the trace of the logarithm seems to be very
complicated. To simplify the situation, we use the trick which was
successfully applied earlier 
\cite{bp2,efpot,BCP}.

Let us consider the free higher-derivative theory of the real scalar superfield whose action is
\bea
S_v=-\frac{1}{16}\int d^8z vD^{\alpha}\bar{D}^2D_{\alpha}\Box v.
\eea
This action is evidently invariant under the usual gauge transformations $\delta v=\Lambda+\bar{\Lambda}$, where $\Lambda$ is a chiral superfield, and $\bar{\Lambda}$ is an antichiral one. Following general prescriptions of the Faddeev-Popov method, one can define the effective action $W_v$ of this theory  as
\bea
\label{efv}
e^{iW_v}=\int Dv \exp(-\frac{i}{16}\int d^8z vD^{\alpha}\bar{D}^2D_{\alpha}\Box v)\delta(\frac{1}{4}D^2v-\bar{\phi})\delta(\frac{1}{4}\bar{D}^2v-\phi),
\eea
where the $\frac{1}{4}D^2v-\bar{\phi}$, $\frac{1}{4}\bar{D}^2v-\phi$
play the role of the gauge fixing functions and $\phi,\bar{\phi}$ are the same as in (\ref{def}). One should notice that
the $W_v$ is a constant independent of $\phi,\bar{\phi}$.

Then, let us multiply the expressions (\ref{def1}) and (\ref{efv}). The functional integration over $\phi,\bar{\phi}$ is straightforward, and we arrive at
\bea
\label{def2}
e^{i\Gamma^{(1)}[\Phi,\bar{\Phi}]+iW_v}=\int Dv \exp(\frac{1}{2}i[\int d^8z( v\Box^2v-v\frac{1}{4}W^{\prime\prime}(\Phi)\bar{D}^2v-v\frac{1}{4}\bar{W}^{\prime\prime}(\bar{\Phi})D^2v)]).
\eea
 The operator whose trace of the logarithm must be calculated to find the one-loop effective action is radically simplified. After omitting irrelevant constants, the one-loop effective action takes the form
\bea
\label{tracelog}
\Gamma^{(1)}=\frac{i}{2}{\rm Tr}\ln (\Box^2-\frac{1}{4}W^{\prime\prime}(\Phi)\bar{D}^2-\frac{1}{4}\bar{W}^{\prime\prime}(\bar{\Phi})D^2).
\eea
Therefore we face the problem of calculating of trace of the logarithm
of the higher-derivative operator.

\section{Superfield proper-time method in the higher-derivative case}

Let us calculate the trace (\ref{tracelog}). The most convenient way for that is based on the use of the Schwinger representation (cf. \cite{efpot}):
\bea
\Gamma^{(1)}&=&\frac{i}{2}{\rm Tr}\int\frac{ds}{s}\exp[is (\Box^2+\frac{1}{4}\Psi\bar{D}^2+\frac{1}{4}\bar{\Psi}D^2)].
\eea
Here we denoted $W^{\prime\prime}(\Phi)=-\Psi$, $\bar{W}^{\prime\prime}(\bar{\Phi})=-\bar{\Psi}$ for the convenience. One should remind that $\Psi$ is a chiral superfield, and $\bar{\Psi}$ is an antichiral one.

Disregarding the terms involving the space-time derivatives of $\Phi,\bar{\Phi}$, which correspond to fourth and higher orders in space-time derivatives of the scalar components of these superfields, we can rewrite this expression as 
\bea
\Gamma^{(1)}&=&\frac{i}{2}\int d^8z\int\frac{ds}{s}\exp[is (\frac{1}{4}\Psi\bar{D}^2+\frac{1}{4}\bar{\Psi}D^2)]e^{is\Box^2}\delta^8(z_1-z_2)|_{z_1=z_2}.
\eea
Now, let us proceed in a way similar to \cite{efpot,bp2}. As a first step, we introduce operators
\bea
\Delta=\frac{1}{4}\Psi\bar{D}^2+\frac{1}{4}\bar{\Psi}D^2;\quad\, \Omega(\Psi,\bar{\Psi},s)=e^{is\Delta}, 
\eea 
where $\Omega$ can be expanded as a power series in the spinor supercovariant derivatives:
\bea
\Omega(\Psi,\bar{\Psi},s)&=&1+\frac{1}{16}A(s)\bar{D}^2D^2+\frac{1}{16}\tilde{A}(s)D^2\bar{D}^2+\frac{1}{8}B^{\alpha}(s)D_{\alpha}\bar{D}^2+\frac{1}{8}\tilde{B}_{\dot{\alpha}}(s)\bar{D}^{\dot{\alpha}}D^2+\nonumber\\&+&\frac{1}{4}C(s)\bar{D}^2+\frac{1}{4}\tilde{C}(s)D^2.
\eea
The $\Omega$ satisfies the superfield heat conductivity equation
\bea
\frac{1}{i}\frac{d\Omega}{ds}=\Omega\Delta.
\eea
The initial condition is evidently $\Omega|_{s=0}=1$, hence $A(s=0)=\tilde{A}(s=0)=B_{\alpha}(s=0)=\tilde{B}_{\dot{\alpha}}(s=0)=C(s=0)=\tilde{C}(s=0)=0$.
The system involving these coefficients turns out to be exactly the same as in the Wess-Zumino case \cite{efpot}, hence the coefficients $A$ and $\tilde{A}$ reproduce the results obtained in that model. 

The one-loop effective action can be expressed as 
\bea
\label{gamma1}
\Gamma^{(1)}&=&-\frac{i}{2}\int d^8z\int\frac{ds}{s}\Omega(\Psi,\bar{\Psi},s)e^{is\Box^2}\delta^8(z_1-z_2)|_{z_1=z_2}.
\eea
Using the well-known properties of the spinor supercovariant derivatives \cite{BK0}, one can show that only the coefficients $A$ and $\tilde{A}$ give nontrivial contributions to the one-loop effective action, i.e.
\bea
\label{gammaa}
\Gamma^{(1)}&=&-\frac{i}{2}\int d^4\theta d^4x_1\int\frac{ds}{s}[A(s)+\tilde{A}(s)]e^{is\Box^2}\delta^4(x_1-x_2)|_{x_1=x_2}.
\eea
The differences with the Wess-Zumino case will arise when, after the expansion of the heat kernel $\Omega(s)$ in series in $\Box$ is carried out, and these d'Alembertians, instead of the usual function $e^{is\Box}\delta^8(z_1-z_2)$, as it occurs in the Wess-Zumino case \cite{efpot}, will act on the function $e^{is\Box^2}\delta^8(z_1-z_2)$. Therefore, at this step, we may merely quote the results obtained for the coefficients $A(s)$, $\tilde{A}(s)$ in the Wess-Zumino case, which, just as in \cite{efpot,BK0}, can be taken up to the fourth order in the spinor supercovariant derivatives of superfields:
\bea
\label{aplus}
A(s)+\tilde{A}(s)&=&\frac{2}{\Box}[\cosh(\tilde{s}U)-1]+\nonumber\\&+&
\tilde{s}\frac{D^2\Psi\bar{D}^2\bar{\Psi}}{64\Box}(\tilde{s}\cosh(\tilde{s}U) -\frac{1}{U}\sinh(\tilde{s}U))+\nonumber\\&+&
\frac{\tilde{s}}{64U^2}[\bar{\Psi}\bar{D}^2\bar{\Psi}(D^{\alpha}\Psi)(D_{\alpha}\Psi)+\Psi D^2\Psi(\bar{D}_{\dot{\alpha}}\bar{\Psi})(\bar{D}^{\dot{\alpha}}\bar{\Psi})]\times\nonumber\\&\times&
(\frac{1}{3}\tilde{s}^2U\sinh(\tilde{s}U)-\tilde{s}\cosh(\tilde{s}U)+\frac{1}{U}\sinh(\tilde{s}U))+\nonumber\\&+&
\frac{\tilde{s}}{256}(D^{\alpha}\Psi)(D_{\alpha}\Psi)(\bar{D}_{\dot{\alpha}}\bar{\Psi})(\bar{D}^{\dot{\alpha}}\bar{\Psi})[\frac{1}{2}\tilde{s}^3\cosh(\tilde{s}U)-\frac{5}{3}\frac{\tilde{s}^2}{U}\sinh(\tilde{s}U)+\nonumber\\&+&
\frac{7}{2U^2}(\tilde{s}\cosh(\tilde{s}U)-\frac{1}{U}\sinh(\tilde{s}U))].
\eea 
Here $\tilde{s}=is$, $U=\sqrt{\Psi\bar{\Psi}\Box}$. The higher orders in supercovariant derivatives of $\Psi$, $\bar{\Psi}$ in principle also can be found. However, obtaining the complete expression for the one-loop superfield potential seems to be an extremely difficult problem.

It remains to substitute these expressions into (\ref{gammaa}) and to expand  (\ref{aplus}) in power series in $\Box$.
The contribution to the one-loop k\"{a}hlerian effective action is given by the first line of (\ref{aplus}), i.e.
\bea
K^{(1)}=-i\int d^4\theta d^4x_1\int\frac{ds}{s}\frac{1}{\Box}[\cosh(\tilde{s}U)-1]e^{is\Box^2}\delta^4(x_1-x_2)|_{x_1=x_2},
\eea
which, after expanding in series in $\Box$ yields
\bea
K^{(1)}=\int d^4\theta d^4x_1\int\frac{dt}{t}\sum\limits_{n=0}^{\infty}\frac{1}{(2n+2)!}(t^2\Psi\bar{\Psi})^{n+1}\Box^n e^{-t\Box^2}\delta^4(x_1-x_2)|_{x_1=x_2}.
\eea
Here we  carried out the Wick rotation $s=it$ (with $t=-\tilde{s}$) and $x_0=ix_{0E}$ for convenience. We also split the indices $n$ into odd, $n=2l+1$ and even, $n=2l$, ones. As a result, this expression takes the form
\bea
\label{dsum}
K^{(1)}&=&\int d^4\theta d^4x_1\int\frac{dt}{t}\sum\limits_{l=0}^{\infty}\left[\frac{1}{(4l+2)!}(t^2\Psi\bar{\Psi})^{2l+1}\Box^{2l}+
\frac{1}{(4l+4)!}(t^2\Psi\bar{\Psi})^{2l+2}\Box^{2l+1}\right]\times\nonumber\\&\times& e^{-t\Box^2}\delta^4(x_1-x_2)|_{x_1=x_2}.
\eea
Now, let us consider the structure $\Box^n e^{-t\Box^2}\delta^4(x_1-x_2)|_{x_1=x_2}$. It is clear that the function $V(t;x_1,x_2)=e^{-t\Box^2}\delta^4(x_1-x_2)$ which we will call the free heat kernel satisfies the equation
\bea
\Box^2V(t;x_1,x_2)=-\frac{d}{dt}V(t;x_1,x_2),
\eea
hence
\bea
\Box^{2l}V(t;x_1,x_2)=(-\frac{d}{dt})^lV(t;x_1,x_2); \quad\,\Box^{2l+1}V(t;x_1,x_2)=(-\frac{d}{dt})^l\Box V(t;x_1,x_2).
\eea
In this paper, the above  expressions will be considered only in the limit $x_1=x_2$. One can find that (cf. \cite{bp2})
\bea
&&V(t;x_1,x_2)|_{x_1=x_2}=\int\frac{d^4k}{(2\pi)^4}e^{-tk^4}=\frac{1}{32\pi^2t};\nonumber\\
&&\Box V(t;x_1,x_2)|_{x_1=x_2}=\int\frac{d^4k}{(2\pi)^4}(-k^2)e^{-tk^4}=-\frac{1}{32\pi^{3/2}t^{3/2}},
\eea
therefore
\bea
\label{boxl}
&&\Box^{2l}V(t;x_1,x_2)|_{x_1=x_2}=(-\frac{d}{dt})^l\frac{1}{32\pi^2t}=\frac{(-1)^ll!}{32\pi^2t^{l+1}};\nonumber\\ &&\Box^{2l+1}V(t;x_1,x_2)|_{x_1=x_2}=(-\frac{d}{dt})^l(-\frac{1}{32\pi^{3/2}t^{3/2}})=-\frac{(-1)^{l+1}(2l+1)!!}{32\pi^{3/2}2^lt^{l+3/2}}.
\eea
Replacing all this into (\ref{dsum}), we arrive at
\bea
\label{dsum1}
K^{(1)}&=&\int d^8z\int\frac{dt}{32\pi^2t}\sum\limits_{l=0}^{\infty}\left[\frac{(t^2\Psi\bar{\Psi})^{2l+1}}{(4l+2)!}\frac{(-1)^ll!}{t^{l+1}}-
\frac{(t^2\Psi\bar{\Psi})^{2l+2}}{(4l+4)!}\frac{(-1)^l\sqrt{\pi}(2l+1)!!}{2^lt^{l+3/2}}\right].
\eea
The series are evidently convergent (to show this, it is sufficient to remind that $\frac{l!}{(4l)!}\leq\frac{1}{(3l)!}$). The integrals over $t$ are also convergent.
An equivalent form of this expression is therefore
\bea
\label{dsum2}
K^{(1)}&=&\int d^8z\int\frac{dt}{32\pi^2t}\sum\limits_{l=0}^{\infty}(-1)^l\left[t^{3l+1}\frac{l!(\Psi\bar{\Psi})^{2l+1}}{(4l+2)!}-
t^{3l+5/2}\frac{(\Psi\bar{\Psi})^{2l+2}}{(4l+4)!}\frac{\sqrt{\pi}(2l+1)!!}{2^l}\right].
\eea
To simplify this expression, let us make the change $t(\Psi\bar{\Psi})^{2/3}=u$ (note that $u$ is dimensionless). We find
\bea
\label{dsum3}
K^{(1)}&=&\int d^8z(\Psi\bar{\Psi})^{1/3}\int\frac{du}{32\pi^2u}\sum\limits_{l=0}^{\infty}\left[\frac{(-1)^lu^{3l+1}l!}{(4l+2)!}-
\frac{(-1)^lu^{3l+5/2}}{(4l+4)!}\frac{\sqrt{\pi}(2l+1)!!}{2^l}\right],
\eea
which can be presented as
\bea
K^{(1)}&=&\frac{c_0}{32\pi^2}\int d^8z(\Psi\bar{\Psi})^{1/3},
\eea
where
\bea
c_0=\int du\sum\limits_{l=0}^{\infty}\left[\frac{(-1)^lu^{3l}l!}{(4l+2)!}-
\frac{(-1)^lu^{3l+3/2}}{(4l+4)!}\frac{\sqrt{\pi}(2l+1)!!}{2^l}\right]
\eea
is a finite constant. It is easy to see that the result for dilaton supergravity \cite{bp2}, being a particular case of this result, is easily reproduced.

Now, let us calculate the one-loop auxiliary fields' effective action. To do it, let us consider all derivative dependent terms in (\ref{aplus}). After their expansion in power series in $\Box$, we find
\bea
\label{f1}
F^{(1)}&=&-i\int d^4\theta d^4x_1\int\frac{dt}{t}\sum\limits_{n=0}^{\infty}\Big[\frac{D^2\Psi\bar{D}^2\bar{\Psi}}{64}t^{2n+4}(\Psi\bar{\Psi})^{n+1}[\frac{1}{(2n+2)!}-\frac{1}{(2n+3)!}]+\nonumber\\&+&\frac{1}{64}[\bar{\Psi}\bar{D}^2\bar{\Psi}D^{\alpha}\Psi D_{\alpha}\Psi+h.c.]
t^{2n+4}(\Psi\bar{\Psi})^n[\frac{1}{3(2n+1)!}-\frac{1}{(2n+2)!}+\frac{1}{(2n+3)!}]+\nonumber\\&+&
\frac{1}{256}D^{\alpha}\Psi D_{\alpha}\Psi\bar{D}_{\dot{\alpha}}\bar{\Psi}\bar{D}^{\dot{\alpha}}\bar{\Psi}t^{2n+6}(\Psi\bar{\Psi})^n
\times\nonumber\\&\times&[\frac{1}{2(2n)!}-\frac{5}{3(2n+1)!}+\frac{7}{2(2n+2)!}-\frac{7}{2(2n+3)!}]
\Big]\times\nonumber\\&\times&\Box^n
e^{-t\Box^2}\delta^4(x_1-x_2)|_{x_1=x_2}.
\eea
Then, we apply the same scheme as above. By its essence, this expression looks like
\bea
F^{(1)}=i\int d^4\theta d^4x_1\int\frac{dt}{t}\sum\limits_{n=0}^{\infty} A_n(\Psi,\bar{\Psi},t)\Box^ne^{-t\Box^2}\delta^4(x_1-x_2)|_{x_1=x_2}.
\eea
Here $A_n$ are some functions of fields whose explicit form can be
read off from (\ref{f1}).
Dividing this sum into sums over odd $n=2l+1$ and even $n=2l$, and taking into account (\ref{boxl}), we find
\bea
F^{(1)}=i\int d^8z \int\frac{dt}{t}\sum\limits_{l=0}^{\infty}[A_{2l}\frac{l!}{32\pi^2t^{2l+1}}-A_{2l+1}\frac{(2l+1)!!}{32\pi^{3/2}2^lt^{2l+3/2}}].
\eea
After carrying out the transformations we used above, we find
\bea
F^{(1)}&=&C_1\frac{\bar{D}^2\bar{\Psi}D^2\Psi}{\Psi\bar{\Psi}}+C_2[\bar{\Psi}\bar{D}^2\bar{\Psi}D^{\alpha}\Psi D_{\alpha}\Psi+h.c.]\frac{1}{(\Psi\bar{\Psi})^2}+\nonumber\\&+&
C_3D^{\alpha}\Psi D_{\alpha}\Psi\bar{D}_{\dot{\alpha}}\bar{\Psi}\bar{D}^{\dot{\alpha}}\bar{\Psi}\frac{1}{(\Psi\bar{\Psi})^2},
\eea
where $C_1,C_2,C_3$ are some numbers, whose explicit form is
\bea
C_1&=&\frac{1}{2048\pi^2}\int du\sum_{l=0}^{\infty}[(\frac{l!}{(4l+2)!}-\frac{l!}{(4l+3)!})u^{3l+2}-\frac{\sqrt{\pi}}{2^l}(\frac{(2l+1)!!}{(4l+4)!}-\frac{(2l+1)!!}{(4l+5)!})u^{3l+7/2}];\nonumber\\
C_2&=&\frac{1}{2048\pi^2}\int du\sum_{l=0}^{\infty}[(\frac{l!}{3(4l+1)!}-\frac{l!}{(4l+2)!}+\frac{l!}{(4l+3)!})u^{3l+2}-\nonumber\\&-&\frac{\sqrt{\pi}}{2^l}(\frac{(2l+1)!!}{(4l+4)!}-\frac{(2l+1)!!}{(4l+5)!})u^{3l+7/2}];
\eea
and
\bea
C_3&=&\frac{1}{8192\pi^2}\int du\sum_{l=0}^{\infty}[(\frac{l!}{2(4l)!}-\frac{5l!}{3(4l+1)!}+\frac{7l!}{2(4l+2)!}-\frac{7l!}{2(4l+3)!})u^{3l+2}-\nonumber\\&-&\frac{\sqrt{\pi}}{2^l}(\frac{(2l+1)!!}{3(4l+5)!}-\frac{(2l+1)!!}{(4l+6)!}+\frac{(2l+1)!!}{(4l+7)!})u^{3l+7/2}].
\eea

To close the consideration of the one-loop effective action for this
model, let us find the one-loop chiral contributions to the effective
action. It is clear that they differ from zero only if  
$\bar{W}^{\prime\prime}(\bar{\Phi})|_{\bar{\Phi}=0}=const\neq 0$
(i.e. if $\bar{\Psi}|_{\bar{\Phi}=0}\equiv\lambda=const$, essentially it means that $\lambda$ is related with the mass of the theory; one should notice that the case $\lambda=0$ gives zero one-loop chiral corrections). We can follow the methodology of \cite{bp2} which, after solving the equations for the coefficients $A,\tilde{A},B_{\alpha},\tilde{B}_{\dot{\alpha}},C,\tilde{C}$ and calculating the traces via the same approach as above yields
\begin{eqnarray}
\label{chir}
{\cal L}^{(1)}_c&=&\lambda^{1/3}
[ \{ (c_1+3c_3) \lambda^{1/3} \Psi^{-1/3} +
c_2 \lambda^{-2/3}\Psi^{2/3} +3c_4 
\lambda^{4/3}\Psi^{-4/3} \}\times\\
&\times&\frac{1}{9}\Psi^{-2}\partial^m\Psi \partial_m\Psi+
\frac{1}{3}(\Psi^{-1}\Box\Psi-\Psi^{-2}\partial^m\Psi \partial_m\Psi) (c_3 \lambda^{1/3}\Psi^{-1/3} +
c_4 \lambda^{4/3}\Psi^{-4/3} )]\nonumber
\end{eqnarray}

The constants $c_1$, $c_2$, $c_3$, $c_4$ have the
form
\begin{eqnarray}
\label{kons}
c_1&=&18\int_0^{\infty}du \sum_{k=0}^{\infty}u^{3k+1} 8^3
\big(\frac{1}{(4k+2)!}-\frac{3}{(4k+3)!}\big)
\frac{{(-1)}^k k!}{32\pi^2} \nonumber\\
c_2&=&-18\int_0^{\infty}du \sum_{k=0}^{\infty}u^{3k+5/2}
\big(\frac{1}{(4k+4)!}-\frac{3}{(4k+5)!}\big)
\frac{{(-1)}^k (2k+1)!!}{2^k 32\pi^{3/2}}\\
c_3&=&6\int_0^{\infty}du\sum_{k=0}^{\infty}u^{3k+1}\frac{1}{(4k+3)!}
\frac{{(-1)}^k k!}{32\pi^2} \nonumber\\
c_4&=&-6\int_0^{\infty}du \sum_{k=0}^{\infty}u^{3k+5/2}
\frac{1}{(4k+5)!}\frac{{(-1)}^k (2k+1)!!}{2^k 32\pi^{3/2}}\nonumber
\end{eqnarray}
All these integrals over $u$ are finite.

We close this section with the conclusion that we have found the lower contributions to the one-loop effective action, involving up to four derivatives.
Now, after we have calculated these contributions, let us study a slightly
 different form of the theory.

\section{Higher-derivatives theory with nonchiral mass term}

In \cite{Ant} the relation between the higher-derivative theory and the usual theory with extended number of superfields was discussed. It was claimed in that paper that,  at the tree level, the theory of a chiral superfield whose kinetic term involves a linear combination of the higher-derivative term $\Phi\Box\bar{\Phi}$  and  the usual one $\Phi\bar{\Phi}$ can be shown to be dynamically equivalent to the theory without higher derivatives but with additional chiral fields. However, it follows from studies of \cite{Ant}, and it can be straightforwardly verified, that this equivalence cannot be established for the theory of the form (\ref{firstex}) which does not contain the usual kinetic term $\Phi\bar{\Phi}$ besides of the higher-derivative one. The studies carried out in the paper \cite{Ant} are applicable only for the theory whose kinetic term is
\bea
S_K=\int d^8z \Phi(\Box-M^2)\bar{\Phi}.
\eea
This kinetic term is equivalent to the one of the Wess-Zumino model
with a higher-derivative regulator \cite{Ili}.
Alternatively, the higher derivatives can be introduced to the superpotential term (we will carry out this analysis elsewhere).
However, the analysis of the effective action for such  theory is more complicated than for the theory studied above. The calculation of the Schwinger coefficients $A(s)$ and $\tilde{A}(s)$ does not differ    from the previous section.   The analog of the free heat kernel function $V(t;x_1,x_2)$, after introducing of the same trick as above, can be shown to be equal to  
\bea
\label{kern}
V(s;x_1,x_2)=e^{-s(\Box^2-M^2\Box)}\delta^4(x_1-x_2).
\eea
However, even the evaluation of the  case $x_1=x_2$, which is only interesting for us in the one-loop approximation, is a nontrivial problem which can be reasonably solved only for very large mass $M$. Let us proceed with this calculation.

After Fourier transform and Wick rotation, the function $V(t;x_1,x_2)|_{x_1=x_2}$ looks like
\bea
I(s)\equiv V(s;x_1,x_2)|_{x_1=x_2}=\int\frac{d^4k}{(2\pi)^4}e^{-s(k^4+k^2M^2)}.
\eea
Changing variables,  $k^2=u$, we find
\bea
I(s)=\frac{1}{16\pi^2}e^{\frac{tM^4}{4}}\int_0^{\infty} duue^{-s(u+\frac{M^2}{2})^2}.
\eea
Replacing  then $u+\frac{M^2}{2}=u'$ and integrating over $u$ where it is possible, we find
\bea
I(s)=\frac{1}{32\pi^2s}-\frac{M^2}{32\pi^2}e^{\frac{sM^4}{4}}\int_{M^2/2}^{\infty}due^{-su^2}.
\eea
We find that this expression for the heat kernel function can be
expressed through the probability integral function
\bea
\Phi(x)=\frac{2}{\sqrt{\pi}}\int_0^x dt e^{-t^2}.
\eea
The presence of such a function seems to make impossible finding  the explicit one-loop k\"{a}hlerian potential in the general case.
It is clear that $\Phi(x\to\infty)\to 1$.
Indeed,
\bea
I(s)=\frac{1}{16\pi^2}(\frac{1}{2s}-\frac{M^2}{2}e^{sM^2/4}(\int_0^{\infty}due^{-su^2}-\int_0^{M^2/2}due^{-su^2}))
\eea
Substituting $su^2=w^2$, we get
\bea
I(s)&=&\frac{1}{16\pi^2}(\frac{1}{2s}-\frac{M^2}{2}e^{sM^2/4}(\frac{1}{2}\sqrt{\frac{\pi}{s}}-\frac{1}{\sqrt{s}}\int_0^{M^2\sqrt{s}/2}dwe^{-w^2}))=
\nonumber\\&=&
\frac{1}{16\pi^2}[\frac{1}{2s}-\frac{M^2}{2}e^{sM^2/4}\frac{1}{2}\sqrt{\frac{\pi}{s}}(1-\Phi(M^2\sqrt{s}/2)].
\eea
To evaluate this expression, we employ the asymptotics of the probability integral $\Phi(y)$ at large arguments \cite{Abr}:
\bea
\Phi(y)|_{y\to\infty}=1-\frac{1}{\pi}e^{-y^2}\sum\limits_{k=0}^{\infty}\frac{(-1)^k\Gamma(k+\frac{1}{2})}{y^{2k+1}}.
\eea
We find that the term with $k=0$ identically cancels the "usual" term $\frac{1}{2s}$. Taking into account only the $M\to\infty$ dominant term (remind that the limit of very high masses was studied earlier in \cite{BCP}), one finds
\bea
I(s)=\frac{1}{16\pi^2s^2M^4},
\eea 
which  differs from the case $M=0$ considered in the
previous section where the analog of this function was proportional to
$\frac{1}{s}$. One could note that such behaviour of the heat kernel seems to be similar to that one occurring in the Wess-Zumino model \cite{efpot}. Nevertheless, the presence of a large mass in the denominator gives a hope that the corrections to the effective action will be suppressed in a $M\to\infty$ limit.

 For simplicity from now on,  we restrict ourselves only to calculation of the k\"{a}hlerian effective potential.

The theory we study here has the action
\bea
\label{twoex}
S[\Phi,\bar{\Phi}]=\int d^8z \Phi(\Box-M^2)\bar{\Phi}+(\int d^6z W(\Phi)+h.c.).
\eea
Here $M$ ia a large parameter related to the physical mass.
Using the insertion of the effective action of the free real scalar superfield whose classical action looks like
\bea
S_v=-\frac{1}{16}\int d^8z vD^{\alpha}\bar{D}^2D_{\alpha}(\Box-M^2) v,
\eea
one can show that the one-loop effective action corresponding to the theory (\ref{twoex}) can be expressed through the following Schwinger representation
\bea
\Gamma^{(1)}&=&\frac{i}{2}{\rm Tr}\int\frac{ds}{s}\exp[is (\Box(\Box-M^2)+\frac{1}{4}\Psi\bar{D}^2+\frac{1}{4}\bar{\Psi}D^2)].
\eea
Since we restrict ourselves here to the k\"{a}hlerian part of the effective potential, we can express the one-loop effective action as
\bea
\Gamma^{(1)}&=&\frac{i}{2}{\rm Tr}\int d^8z\int\frac{ds}{s}
\exp[is (\frac{1}{4}\Psi\bar{D}^2+\frac{1}{4}\bar{\Psi}D^2)]e^{is\Box(\Box-M^2)}\delta^8(z-z')|_{z=z'}.
\eea
The relevant terms from the operator $\exp(is (\frac{1}{4}\Psi\bar{D}^2+\frac{1}{4}\bar{\Psi}D^2))$ again have the form (\ref{aplus}),
and the one-loop k\"{a}hlerian effective action looks like
\bea
K^{(1)}=-i\int d^4\theta d^4x_1\int\frac{dt}{t}\frac{1}{\Box}[\cosh(t\sqrt{\Psi\bar{\Psi}\Box})-1]e^{-t\Box(\Box-M^2)}\delta^4(x_1-x_2)|_{x_1=x_2}.
\eea
Expanding this in series in $\Box$, after Wick rotation we find
\bea
K^{(1)}=\int d^4\theta d^4x_1\int\frac{dt}{t}\sum\limits_{n=0}^{\infty}\frac{1}{(2n+2)!}(t^2\Psi\bar{\Psi})^{n+1}\Box^n V(t;x_1,x_2)|_{x_1=x_2}.
\eea
Here the function $V(t;x_1,x_2)$ can be read off from the (\ref{kern}).
As we already noted, this expression can be found in a closed form
only 
in the limit $M\to\infty$.
It follows from (\ref{kern}) that
\bea
\Box^n V(t;x_1,x_2)=\frac{1}{t^n}(\frac{d}{d(M^2)})^nV(t;x_1,x_2),
\eea
so that, after taking $x_1=x_2$ , 
\bea
\Box^n V(s;x_1,x_2)|_{x_1=x_2}=\frac{(-1)^n(n+1)!}{16\pi^2(M^2t)^{n+2}}.
\eea
Putting all together, we find
\bea
K^{(1)}=\frac{1}{32\pi^2}\int d^8z\int\frac{dt}{M^2t^2}\sum\limits_{n=0}^{\infty}\frac{(-1)^n(n+1)!}{(2n+2)!}\left(\frac{t^2\Psi\bar{\Psi}}{M^2}\right)^{n+1}.
\eea
This expression is similar to that one obtained in \cite{efpot} for the Wess-Zumino model. As a result, we have
\bea
K^{(1)}=\frac{1}{32\pi^2}\int d^8z \frac{\Psi\bar{\Psi}}{M^4}\sum\limits_{n=0}^{\infty}\int_{\frac{\Psi\bar{\Psi}L^2}{M^4}}^{\infty}\frac{du}{u}\frac{(-1)^nu^n(n+1)!}{(2n+2)!}.
\eea
To avoid divergence of the integral,
 we introduced the cutoff $L^2$ at the lower limit. As $L^2\to 0$, one obtains
\bea
K^{(1)}&=&-\frac{1}{32\pi^2}\frac{\Psi\bar{\Psi}}{M^4}\ln(\mu^2L^2)-\frac{1}{32\pi^2}\frac{\Psi\bar{\Psi}}{M^4}(\ln\frac{\Psi\bar{\Psi}}{M^4\mu^2}-\xi).
\eea
Here $ \xi $ is some finite constant which can be absorbed into a redefinition of $ \mu^2 $.
This contribution is divergent but turns out to be suppressed in the large $M$ limit.

\section{Summary}

We considered the one-loop effective potential for different versions of
the  higher-derivative chiral superfield models. 
It turns out  that, in the case when the mass term is purely
chiral (a similar situation with the mass term takes place in the
Wess-Zumino model), the theory is finite. At the same time, if the
mass term arises in the general Lagrangian (that is the situation
considered in \cite{Ant}), the theory displays divergences though
being super-renormalizable. We note, however, that the equivalence of
the higher-derivative theory of the chiral superfield and the theory
without higher derivatives but with an extended number of chiral
superfields described in \cite{Ant} occurs only in the case when the
mass term belongs to the general Lagrangian (that is, the second case
considered in the paper). Therefore, the presence of these divergences can
be considered as a sign in favour of the equivalence established in \cite{Ant}. On the other hand, a detailed study of the effective action in a theory involving several chiral superfields seems to be technically complicated, so we consider it in another work.

{\bf Acknowledgements.} This work was partially supported by Conselho
Nacional de Desenvolvimento Cient\'{\i}fico e Tecnol\'{o}gico (CNPq)
and Funda\c{c}\~{a}o de Amparo \`{a} Pesquisa do Estado de S\~{a}o
Paulo (FAPESP), Coordena\c{c}\~{a}o de Aperfei\c{c}oamento do Pessoal
do Nivel Superior (CAPES: AUX-PE-PROCAD 579/2008) and CNPq/PRONEX/FAPESQ.

\end{document}